\documentstyle[times,graphics,epsfig,amsmath,amssymb,natbib,psfrag,color]{mn2e}

\newcommand{\be}{\begin{equation}}
\newcommand{\e}{\end{equation}}
\newcommand{\bear}{\begin{eqnarray}}
\newcommand{\ear}{\end{eqnarray}}

\newcommand{\R}{{\cal R}}
\def\xh1{x_{H~{\sc i}\,}}
\def\xi{x^{i}_{{\rm H~{\sc{i}}}}\,}
\def\xb{\bar{x}_{{\rm H~{\sc{i}}}}}
\def\Ph1{P_{{H~{\sc{i}}}}}
\def\eh1{\eta_{{H~{\sc{i}}}}}
\def\r{r}
\def\a{{\mathcal A}}
\def\aap{AAP}
\def\apj{ApJ}
\def\aj{AJ}
\def\apjs{ApJS}
\def\apjl{ApJL}
\def\mnras{MNRAS}

\def\araa{Annu. Rev. Astro. Astrophys.}

\def\k{{\bf k}}

\def\HI{H~{\sc i}\,}

\def\P{\hat{P}}
\def\Pl{{\mathcal P}}
\def\k{{\bf k}}
\def\R{{\mathcal R}}
\def\Rh{\hat{\mathcal{R}}}
\begin{document}
\title[EoR 21-cm signal peculiar velocities effect]{The effect of
  peculiar velocities on the epoch of reionization (EoR) 21-cm signal}

\author[Majumdar, Bharadwaj \& Choudhury ]{Suman
  Majumdar$^{1,2}$\thanks{E-mail: sumanm@phy.iitkgp.ernet.in}, Somnath
  Bharadwaj$^1$\thanks{E-mail: somnath@phy.iitkgp.ernet.in} and T. Roy
  Choudhury$^{3}$\thanks{E-mail: tirth@ncra.tifr.res.in} \\
  $^1$Department of Physics and Meteorology \& Centre for Theoretical
  Studies, IIT, Kharagpur 721302, India\\
  $^2$ Department of Astronomy \& Oskar Klein Centre, AlbaNova,
  Stockholm University, SE-106 91 Stockholm, Sweden \\
  $^3$National Centre for Radio Astrophysics, TIFR, Post Bag 3,
  Ganeshkhind, Pune 411007, India}

\maketitle

\begin{abstract}
  We have used semi-numerical simulations of reionization to study the
  behaviour of the power spectrum of the EoR 21-cm signal in redshift
  space. We have considered two models of reionization, one which has
  homogeneous recombination (HR) and the other incorporating
  inhomogeneous recombination (IR). We have estimated the observable
  quantities --- quadrupole and monopole moments of \HI power spectrum
  at redshift space from our simulated data. We find that the
  magnitude and nature of the ratio between the quadrupole and
  monopole moments of the power spectrum ($P^s_2 /P^s_0$) can be a
  possible probe for the epoch of reionization. We observe that this
  ratio becomes negative at large scales for $\xb \leq 0.7$
  irrespective of the reionization model, which is a direct signature
  of an inside-out reionization at large scales. It is possible to
  qualitatively interpret the results of the simulations in terms of
  the fluctuations in the matter distribution and the fluctuations in
  the neutral fraction which have power spectra and cross-correlation
  $P_{\Delta \Delta}(k)$, $P_{xx}(k)$ and $P_{\Delta x}(k)$
  respectively. We find that at large scales the fluctuations in
  matter density and neutral fraction is exactly anti-correlated
  through all stages of reionization. This provides a simple picture
  where we are able to qualitatively interpret the behaviour of the
  redshift space power spectra at large scales with varying $\xb$
  entirely in terms of a just two quantities, namely $\xb$ and the
  ratio $P_{xx}/P_{\Delta \Delta}$. The nature of $P_{\Delta x}$
  becomes different for HR and IR scenarios at intermediate and small
  scales. We further find that it is possible to distinguish between
  an inside-out and an outside-in reionization scenario from the
  nature of the ratio $P^s_2 /P^s_0$ at intermediate length
  scales. 
\end{abstract}


\section{Introduction}
The epoch when the neutral hydrogen (\HI) in the inter-galactic medium
(IGM) was reionized by the first luminous sources, is one of the least
known periods in the history of our universe. Observations of the CMBR
\citep{spergel03,page07,komatsu11,larson11} and absorption spectra of
high redshift quasars
\citep{becker01,fan03,white03,fan06,willott07,goto11} suggest that the
epoch of reionization (EoR) probably extended over the redshift range
$6 \leq z \leq 15$
\citep{fan06,choudhury06a,alvarez06,mitra11}. However these
observations are limited in their ability to shed light on many
important questions regarding EoR. What are the major sources of
reionization? What are the typical sizes and the topology of the
ionized regions at different stages?  Observations of redshifted 21-cm
radiation from neutral hydrogen hold the promise to answer some of
these questions.  The brightness temperature of the redshifted 21-cm
radiation directly probes the \HI distribution at the epoch where the
radiation originated.  It is thus possible to track the entire
reionization history as it gradually proceeds with redshift. The
presently functioning low frequency radio telescopes
{GMRT\footnote{http://www.gmrt.ncra.tifr.res.in}} \citep{swarup}, 
LOFAR\footnote{http://www.lofar.org/} and 
{21CMA\footnote{http://21cma.bao.ac.cn/}}, the upcoming
{MWA\footnote{http://www.haystack.mit.edu/ast/arrays/mwa/}} and the
future {SKA\footnote{http://www.skatelescope.org/}} all cover the
frequency range relevant for the EoR 21-cm signal, and this is one of
the major goals for most of these telescopes. It is therefore very
important to have a good picture of the expected signal in order to
make forecasts for and correctly interpret the future observations of
the redshifted 21-cm radiation.

There has been a considerable amount of work towards simulating the
expected EoR 21-cm signal. In particular, there have been numerical
simulations which use ray-tracing to follow the propagation of
ionization fronts in the IGM
\citep{gnedin00,ciardi01,ricotti02,razoumov02,maselli03,sokasian03,iliev06,mellema06,mcquinn07,trac07,semelin07,shin08,iliev08,shapiro08,thomas09,baek09}.
Such simulations are computationally extremely expensive, and it is
difficult to simulate large volumes, and to re-run the simulations
considering different values of the simulation
parameters. Semi-numerical simulations which consider the average
photon density in place of a detailed ray-tracing analysis provide a
computationally less expensive technique to simulate the EoR 21-cm
signal
\citep{furlanetto04,mesinger07,geil08a,lidz09,choudhury09,alvarez09,santos10,mesinger11,zahn11}.

The fluctuations in the brightness temperature of the redshifted 21-cm
radiation essentially trace the \HI distribution during EoR.  The
redshift space distortion caused by peculiar velocities also plays an
important role in shaping the redshifted 21-cm signal
\citep{bharadwaj01,bharadwaj04}. In fact, we expect the peculiar
velocities to introduce an anisotropy in the three dimensional power
spectrum of the EoR 21-cm signal \citep{barkana05,bharadwaj05,wang06},
very similar to the characteristic anisotropy present in the galaxy
power spectrum \citep{kaiser87}. \citet{barkana05} have proposed that
it may be possible to use this anisotropy to separate the effect of
the peculiar velocities from the other astrophysical information
present in the 21-cm power spectrum.

Until recently, most simulations of the EoR 21-cm signal have not
considered the effect of redshift space distortions. Some of the
earlier work \citet{mellema06} and \citet{thomas09} have considered
this effect while generating \HI maps through their simulations, but
have not studied it's implication on the statistical properties like
the power spectrum of the brightness temperature
fluctuations. Recently, \citet{santos10} and \citet{mesinger11} have
included the effect of redshift space distortions in an approximate,
perturbative fashion in their semi-numerical simulation and used this
to study it's implications on the redshifted 21-cm brightness
temperature power spectrum at different stages of the reionization.
In a very recent work \citet{mao12} discuss the methodology to
implement redshift space distortion in numerical simulations of
reionization, and used this to study the 21-cm brightness temperature
power spectrum during EoR.

Most of the earlier semi-numerical simulations ({\it e.g.}
\citealt{furlanetto04,mesinger07,geil08a,lidz09,alvarez09,santos10,mesinger11,zahn11})
have assumed spatially homogeneous recombination which predicts
strictly inside-out reionization where the most dense regions ionize
first, the ionization subsequently propagating to lower densities.
However, there are observations which indicate exactly opposite
picture at the end of reionization, where the high density regions
remain neutral (due to self-shielding) and the low density regions are
highly ionized. \citet{choudhury09} have attempted to make their
semi-numerical simulation consistent with these observations by
incorporating the fact that recombination occurs faster in high
density regions. In these simulations reionization is inside-out only
in the early stages. However, self-shielded, high density clumps
remain neutral in the later stages of reionization when inhomogeneous
recombination is taken into account. In this paper we follow
\citet{choudhury09} to develop a semi-numerical code to simulate
reionization, with the further improvement that we incorporate the
effect of redshift space distortion due to peculiar velocities.  We
have used these simulations to study the effect of peculiar velocities
on the EoR 21-cm signal, both with homogeneous recombination and with
inhomogeneous recombination.

In this paper we have used semi-numerical simulations to determine the
EoR 21-cm signal at different stages of reionization, and used the \HI
power spectrum to quantify the statistical properties of this
signal. We have calculated $P^r_{\HI}(k)$ the \HI power spectrum in
real space and its redshift space counterpart $P^s_{\HI}({\bf k})$,
and compared these two to asses the effect of peculiar velocities. The
anisotropy of the 21-cm signal, quantified through various angular
multipoles of $P^s_{\HI}({\bf k})$, is a very useful tool to study the
effect of redshift space distortion.  In particular, we have studied
the monopole and quadrupole moments of $P^s_{\HI}({\bf k})$ in order
to identify the features characteristic of redshift space distortion
at different stages of reionization. To our knowledge, this anisotropy
has not been quantified using simulations in any of the earlier
studies.  Finally, we attempt to interpret the results of our
simulations, and compare these against the predictions of the simple,
linear model proposed by \citet{barkana05}.

Unless mentioned otherwise, throughout this paper we present results
for the cosmological parameters $h= 0.704$, $\Omega_m = 0.272$,
$\Omega_{\Lambda} = 0.728$, $\Omega_b h^2 = 0.0226$ (all parameters
from WMAP 7 year data \citep{komatsu11,jarosik11}).

A brief summary of the paper follows. In Section 2.  we present the
semi-numerical technique that we have used to simulate the EoR 21-cm
signal including the effect of peculiar velocities. Section
3. contains a brief discussion of the model prediction for the effect
of redshift space distortion. These were used as reference values in
presenting and interpreting the results from our simulations in
Section 4.  Finally, we discuss our results and conclude in Section 5.

\section{Simulating redshift space distortion during Reionization}
\label{sec:sim}
We have used a semi-numerical simulation to generate the \HI
ionization map during reionization. The simulation essentially starts
from the dark matter distribution at a given redshift, and uses this
to identify the sources of ionizing photons. These sources, along with
the assumption that the \HI traces the dark matter are used to
construct a snapshot of the \HI ionization distribution. Our
simulation is based on the formalism proposed by
\citet{choudhury09}. This uses an excursion-set formalism as
introduced by \citet{furlanetto04}. The semi-numerical simulation
provides us with the ionization field in the real space {\it i.e.}
without the redshift space distortion. We briefly discuss the
semi-numerical method that we have used in this work to simulate the
brightness temperature fluctuations of the redshifted 21-cm emission
from EoR.

We have used a Particle Mesh N-body code to generate the dark matter
distribution. The spatial and mass resolution of the N-body simulation
should be adequate to correctly resolve all the ionizing sources that
one is going to adopt in this semi-numerical simulation. It is
currently believed that the stars residing in the galaxies are the
major source of photons to reionize the universe
\citep{yan04,stiavelli04,bouwens05,fan06,choudhury06}. The presently
accepted models suggest that dark matter halos having a mass $M \geq
10^9 \,h^{-1}\, M_{\odot}$ host the early galaxies that contribute to
reionization.  We thus include all dark matter halos of mass $M \geq
10^9 \,h^{-1}\, M_{\odot}$ in our semi-numerical simulation.  Assuming
that at least $10$ dark matter particles are required to constitute
the smallest halo, the N-body simulation is required to have a mass
resolution $\leq 10^8 \,h^{-1}\, M_{\odot}$.

We have generated the dark matter distribution at $z = 8$ using the 
Particle Mesh N-body code. The volume of the simulation is constrained
by the $16$ Gigabytes of memory available in our computer. We perform
our simulation in a periodic box of size $85.12$ Mpc (comoving) with
$1216^3$ grid points and $608^3$ particles, with a mass resolution
$M_{part} = 7.275 \times 10^7\,h^{-1}\,M_{\odot}$.

We identify halos within the simulation box using a standard
Friend-of-Friend algorithm \citep{davis}, with a fixed linking length
$0.2$ (in units of mean inter particle distance) and minimum dark
matter halo mass $ = 10 M_{part}$.

The relation between the ionizing luminosity of a galaxy and its
properties is not well known from the observations. In the
semi-numerical formalism adopted here, we assume that the ionizing
luminosity from a galaxy is proportional to the mass of it's halo. The
number of ionizing photons contributed by a halo of mass $M$ is given
by
\begin{equation}
N_{\gamma}(M) = N_{ion} \frac{M}{m_H}
\label{eq:nion} 
\end{equation}
where $m_H$ is the mass of a hydrogen atom and $N_{ion}$ is a
dimensionless constant. The value of $N_{ion}$ is tuned so as to
achieve the desired mass-averaged neutral fraction   in the
simulation. The ionizing 
photon field is estimated on a grid which has a resolution $8$ times
coarser than  the N-body simulation.

In our simulation we assume that the baryons follow the dark matter
distribution and we also assume that each N-body particle has same
hydrogen mass $M_H$. In the semi-numerical formalism, a region is said
to be ionized if the average number of photons reaching there exceeds
the average neutral hydrogen density at that point. Before applying
the ionization condition we assume that the entire hydrogen contained
in each particle is completely neutral. Using this assumption we
calculate the \HI density field in the same grid where the photon
field has been generated. The photon and the \HI density at each grid
point are compared and a neutral fraction $\xh1$ is assigned to the
grid point depending on the ionization conditions discussed below.

In this work we consider two different models of reionization. In one
model the recombination rate is assumed to be homogeneous (HR) and
independent of density throughout the IGM.  In the other model we have
considered a density dependent inhomogeneous recombination
(IR). Readers are requested to refer to the Section 2 of
\citet{choudhury09} for further details of ionization conditions
(eq. [7] and [15] of \citealt{choudhury09}) in these two different
models of reionization. For the simulation results related to IR model
presented in this paper we have set the inhomogeneous recombination
parameter $\epsilon = 1.0$ (see Section 2.5 and eq. [15] of
\citealt{choudhury09} for the definition of $\epsilon$).

We next discuss how we implement the effect of peculiar velocities on
the ionization maps generated.  We first consider the dark matter
particles to each of which we have assigned a total hydrogen mass
$M_{H}$.  The ionization map provides us with a neutral fraction
$\xh1$ at each grid point of the simulation.  For the $i$th particle
in the simulation, we have interpolated the neutral fraction from its
eight nearest neighbouring grid points to determine the neutral
fraction $x^i_{\HI}$ at the particle's position.  We use this to
calculate the particle \HI mass $M^i_{\HI}$ as
\begin{equation}
M^i_{\HI} = \xh1^i M_{H}\, .
\end{equation}
This provides us with the \HI distribution and the peculiar velocity
associated with each \HI element. We now consider a distant observer
located along the $x$ axis, and use the $x$ component of the peculiar
velocity to determine the particle positions in redshift space  
\begin{equation}
s = x + \frac{v_{x}}{a H(a)}
\label{eq:rsd}
\end{equation}
where $a$ and $H(a)$ are the scale factor and Hubble parameter
respectively. Finally we have interpolated the \HI distribution from
the particles to the grid, and used this to generate the EoR 21-cm
signal. This method of mapping the real space \HI density in redshift
space is some what similar to the PPM-RRM scheme discussed in
\citet{mao12}.
\section{Modeling redshift space distortion during reionization}
\label{sec:model}
\begin{figure*}
\psfrag{xh1=0.8}[c][c][1][0]{$\xb=0.8$}
\psfrag{xh1=0.7}[c][c][1][0]{$\xb=0.7$}
\psfrag{xh1=0.5}[c][c][1][0]{$\xb=0.5$}
\psfrag{xh1=0.3}[c][c][1][0]{$\xb=0.3$} 
\psfrag{HR}[c][c][1][0]{HR}
\psfrag{HR-RS}[c][c][1][0]{HR-RS} 
\psfrag{IR}[c][c][1][0]{IR}
\psfrag{IR-RS}[c][c][1][0]{IR-RS} 
\includegraphics[width=1.\textwidth,
  angle=0]{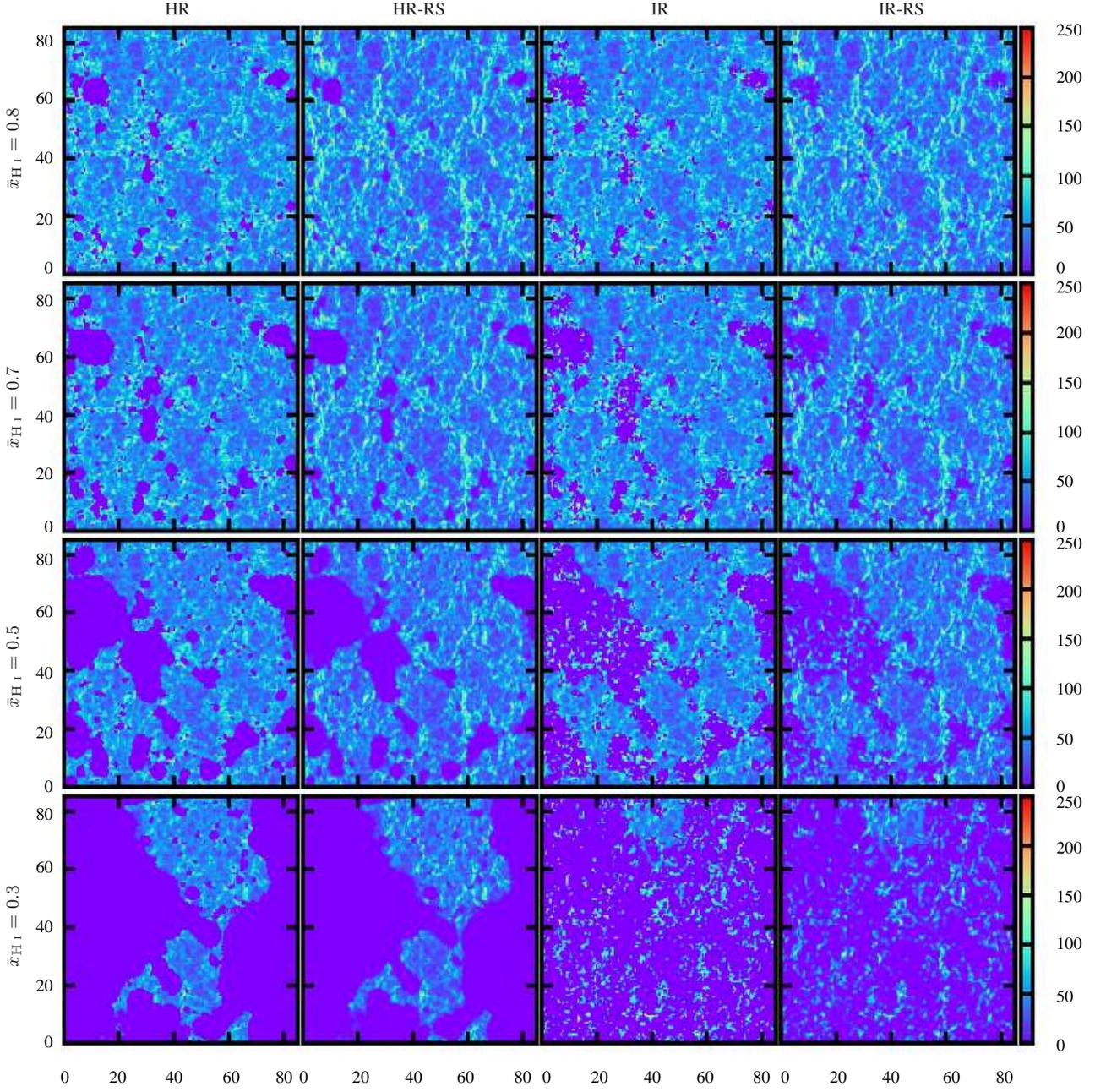}
  \caption{Simulated \HI maps in real and redshift space with
    different schemes of reionizations for various values of $\xb$. In
    this figure we use HR, HR-RS, IR and IR-RS to denote homogeneous
    recombination, homogeneous recombination with redshift space
    distortion, inhomogeneous recombination and inhomogeneous
    recombination with redshift space distortion respectively. The
    line of sight is along the x-axis and the coordinates are in
    Mpc. The colour index represents \HI density in arbitrary units.}
\label{fig:h1_map}
\end{figure*}

Coherent inflows into overdense regions and outflows from underdense
regions  appear as   enhancements in the matter density
fluctuations observed in redshift space.  This introduces an anisotropy
\citep{kaiser87} in $P^s (k,\mu)$ the redshift space matter power
spectrum  
\begin{equation}
P^s ({k,\mu}) = \left( 1 +  \mu^2 \right)^2 P^r(k)
\label{eq:kaiser}
\end{equation}
 where   $P^r(k)$ is the  
real space power spectrum  and  $\mu={\bf k}
\cdot {\bf \hat{n}}/k$ is the cosine of the angle 
between the wave vector ${\bf k}$ and the unit vector ${\bf n}$ along
the line of sight (LoS). Here we have assumed $\Omega_m =1$
throughout, which is 
reasonable at the high redshifts  of our interest. 
It is convenient \citep{hamilton92,hamilton98,cole95} to decompose
the anisotropy using Legendre polynomials $ {\mathcal P}_l(\mu)$ 
as 
\begin{equation}
P^s(k,\mu) = \sum_{l \,{\rm even}} {\mathcal P}_l(\mu)
P_{l}^s(k)\, ,
\end{equation}
where  $P_{l}^s(k)$ are the different angular multipoles of 
$P^s(k,\mu)$.  Under the linear approximation (eq. \ref{eq:kaiser}),
only the first three even moments have non-zero values 
$P_{0}^s(k)/P^r(k) = 28/15 $,
$P_{2}^s(k)/P^r(k) = 40/21 $ and $P_{4}^s(k)/P^r(k) = 8/35$ which are
constant independent of $k$. 

Peculiar velocities have a similar effect on the 
 brightness temperature  fluctuations $\Delta T_b$ of the 21-cm \HI
radiation  from the high redshift universe
\citep{bharadwaj01,bharadwaj04}. Expressing the brightness temperature
fluctuations as $\Delta T_b({\bf x}, z)= \bar{T}(z) \,  \eta_{\HI}({\bf
  x}, z)$ and  considering the EoR where the spin
temperature is much higher than the CMBR temperature 
($T_S \gg T_{\gamma}$)  we have \citep{bharadwaj05} 
\begin{equation}
\eta_{\HI}({\bf x}, z) =  \frac{\rho_{\HI}}{\bar{\rho}_{H}} \left[ 1 -
  \frac{(1+z)}{H(z)}\frac{\partial v_{\parallel}}{\partial r}
  \right]
\label{eq:tb}
\end{equation}
where all the quantities in the r.h.s. refer to the position and
 epoch where the \HI emission originated and 
$ \bar{T}(z) = 4.0\, {\rm mK} (1+z)^2 \left(\frac{\Omega_b
  h^2}{0.02}\right) \left(\frac{0.7}{h}\right)  \frac{H_0}{H(z)} $.
 Here $\rho_{\HI}$ refers to
 the \HI density (which varies from position to position),
 $\bar{\rho}_{H}$ is the mean hydrogen density, $r$ is the comoving
 distance from the observer and $v_{\parallel}$ is the radial
 component of the peculiar velocity. We may express $\rho_{\HI}$ using 
 $\rho_{\HI}/\bar{\rho}_{H}= x_{\HI} (1 + \delta) $ where  $x_{\HI}$
 is the  hydrogen neutral fraction  and $\delta$ is the matter
 overdensity     which  we have assumed to be the same as the  
hydrogen overdensity  at the the large length-scales of our interest.  
Further, we may express the neutral fraction as $x_{\HI}=\langle \xh1
\rangle_{V} (1 + \delta_x)$ where  $\langle \xh1
\rangle_{V}$ in the volume averaged neutral fraction and $\delta_x$ is
the contrast in the $x_{\HI}$ distribution. 
Note that the volume averaged neutral fraction $\langle \xh1
\rangle_{V}$ and  the mass averaged neutral fraction $\bar{x}_{\HI}$ 
refer to the average of $x_{\HI}$ and $x_{\HI}(1+\delta)$
respectively,    
Assuming that $\delta_x,\delta,(\partial v_{\parallel}/\partial r) \ll
1$  we  drop all quadratic and higher terms involving  
$\delta_x,\delta$ and $(\partial v_{\parallel}/\partial r)$,
to obtain $\langle \xh1
\rangle_{V}= \bar{x}_{\HI}$ whereby it is possible 
to express $\eta_{\HI}$ in Fourier space as 
\begin{equation}
\tilde{\eta}_{\HI}({\bf k}) = \bar{x}_{\HI}
\left[ \Delta_x + (1 + \mu^2)\Delta \right] \,.
\label{eq:d_tb}
\end{equation}
where $\tilde{\eta}_{\HI},\Delta$ and $\Delta_x$ are the Fourier
transform of $\eta_{\HI}, \delta$ and $\delta_x$ respectively. 
This gives the redshift space \HI power spectrum  (of $\eta_{\HI}$) to be 
\citep{barkana05} 
\begin{align}
P^s_{\HI} (k,\mu) = \bar{x}^2_{\HI} &\left[P_{xx} (k) + 2 (1 + \mu^2)
  P_{\Delta x} (k)\right.\nonumber\\  &\left. + (1 + \mu^2)^2 P_{\Delta \Delta} (k) \right]
\label{eq:model}
\end{align}
where $P_{xx}$ and $P_{\Delta \Delta}$ are the power spectra of
$\Delta_x$ and $\Delta$ respectively, and $P_{\Delta x}$ is the cross
power spectrum between $\Delta$ and $\Delta_x$. We recover the real
space \HI power spectrum $P^r(k)$ if we set $\mu=0$ in
eq. (\ref{eq:model}). Here, and in the subsequent discussion, we drop
the subscript \HI for brevity of the symbols. Thus the \HI power
spectrum in real space will be
\begin{equation}
P^r  = \bar{x}^2_{\HI} \left( P_{\Delta \Delta} + 2 P_{\Delta x}+ P_{xx} \right)  \,.
\label{eq:modelr}
\end{equation}
In this model only the first three even angular moments of the
redshift space power spectrum have non-zero values
\begin{equation}
  P^s_0  = \bar{x}^2_{\HI} \left( \frac{28}{15} P_{\Delta \Delta} + \frac{8}{3} P_{\Delta x}+
    P_{xx} \right)  \,,
\label{eq:model0}
\end{equation}
\begin{equation}
P^s_2 = \bar{x}^2_{\HI} \left( \frac{40}{21} P_{\Delta \Delta} + \frac{4}{3} P_{\Delta x} \right) \,,
\label{eq:model2}
\end{equation}
\begin{equation}
P^s_{4} = \bar{x}^2_{\HI} \left( \frac{8}{35} \right) P_{\Delta \Delta}\,.
\label{eq:model4}
\end{equation}
To provide an interpretation of our results obtained from
semi-numerical simulations in this paper we have used this linear
model. Finally, we note that this model is based on the assumption
$\delta_x \ll 1$, and terms of the order of $\delta_x \delta$ which
appear in eq. (\ref{eq:tb}) are ignored. While this is possibly a
reasonable assumption in the early stages of reionization
($\bar{x}_{\HI} \sim 1$), we may expect significant deviations from
this model in the late stages of the reionization where
$\bar{x}_{\HI}$ is small and we expect $\delta_x$ to exhibit large
fluctuation of order unity. \citet{lidz07} and \citet{mao12} have
considered models which incorporate the non-linear effects, but these
are rather complicated and we have not attempted using these models
here.

\subsection{Method to estimate the angular multipoles of \HI power
  spectrum from the simulated \HI maps}
We Fourier transform the entire simulated image data cube and estimate
the angular multipoles $P^s_l$ of \HI power spectrum from the Fourier
transformed data following the equation
\begin{equation}
  P^s_{l} (k) = \frac{\left( 2l + 1 \right)}{4 \pi} \int {\mathcal P}_l(\mu)\, P^s (k)\, d\Omega \,,
\label{eq:p_sl}
\end{equation}
where $P^s (k)$ is the \HI power spectrum in the redshift space which
has been estimated from the Fourier transform of the simulated
redshift space \HI map.  The integral is done over the entire solid
angle to take into account all possible orientations of the ${\bf k}$
vector with the LoS direction ${\bf \hat{n}}$.  Each angular multipole
is estimated at $10$ logarithmically spaced $k$ bins in the range
$0.09\leq k \leq 4.80\, {\rm Mpc} ^{-1}$.
   
\section{Results}
\begin{figure*}
\psfrag{xh1=1.0}[c][c][1][0]{$\xb=1.0$}
\psfrag{=0.9}[c][c][1][0]{  $\,\,0.9$}
\psfrag{=0.8}[c][c][1][0]{  $\,\,0.8$}
\psfrag{=0.7}[c][c][1][0]{  $\,\,0.7$}
\psfrag{=0.5}[c][c][1][0]{  $\,\,0.5$}
\psfrag{=0.3}[c][c][1][0]{  $\,\,0.3$}
\psfrag{=0.1}[c][c][1][0]{  $\,\,0.1$}
\psfrag{IR}[c][c][1][0]{{\large IR}}
\psfrag{HR}[c][c][1][0]{{\large HR}}
\psfrag{k}[c][c][1][0]{{\large $k$ in ${\rm Mpc}^{-1}$}}
\psfrag{LT}[c][c][1][0]{LT}
\psfrag{Ps/Pr}[c][c][1][0]{{\large $P^s_0(k)/P^r(k)$}}
\centering
\includegraphics[width=0.55\textwidth, angle=-90]{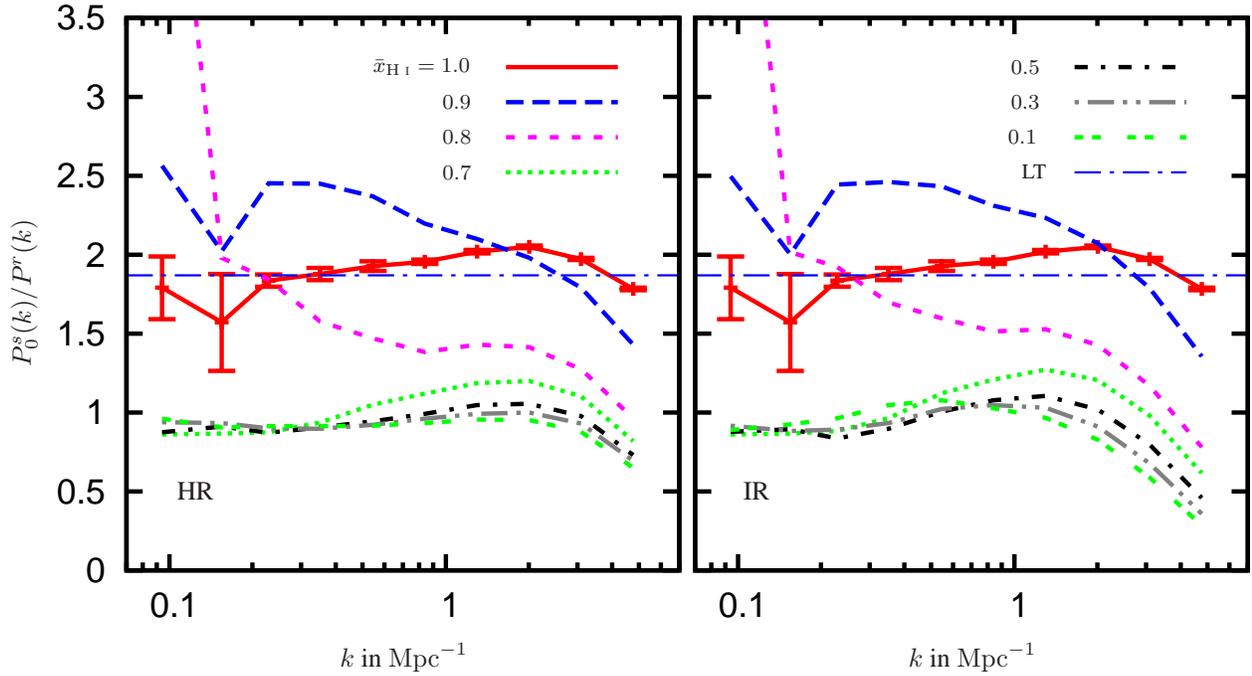}
\caption{The ratio $P^s_0(k)/P^r(k)$ for different neutral fractions
  and for both the models of reionization. For reference, the
  horizontal thin dash-dot line (LT) shows the value $28/15$ predicted
  by the linear theory of redshift space distortion for a completely
  neutral IGM. The error bars represent the $1\sigma$ sample variance
  estimated from $12$ independent realizations.}
\label{fig:pk_ratio}
\end{figure*}
\begin{figure*}
\psfrag{xh1=1.0}[c][c][1][0]{$\xb=1.0$}
\psfrag{=0.9}[c][c][1][0]{ $\,0.9$}
\psfrag{=0.8}[c][c][1][0]{ $\,0.8$}
\psfrag{=0.7}[c][c][1][0]{ $\,0.7$}
\psfrag{=0.5}[c][c][1][0]{ $\,0.5$}
\psfrag{=0.3}[c][c][1][0]{ $\,0.3$}
\psfrag{=0.1}[c][c][1][0]{ $\,0.1$}
\psfrag{IR}[c][c][1][0]{{\large IR}}
\psfrag{HR}[c][c][1][0]{{\large HR}}
\psfrag{LT}[c][c][1][0]{LT}
\psfrag{k}[c][c][1][0]{{\large $k$ in ${\rm Mpc}^{-1}$}}
\psfrag{Ps2/Ps0}[c][c][1][0]{{\large $P^s_2(k)/P^s_0(k)$}}
\centering
\includegraphics[width=.55\textwidth, angle=-90]{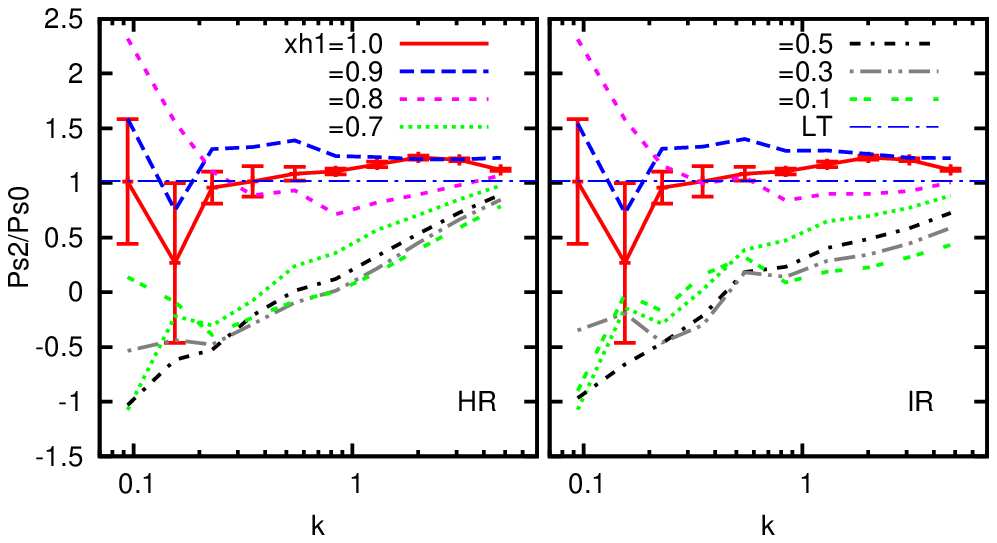}
\caption{The ratio $P^s_2(k)/P^s_0(k)$ for different neutral fractions
  and for both the models of reionization. For reference, the
  horizontal thin dash-dot line (LT) shows the value $50/49$ predicted
  by the linear theory of redshift space distortion for a completely
  neutral IGM. The error bars represent the $1\sigma$ sample variance
  estimated from $12$ independent realizations.}
\label{fig:p2p0_ratio}
\end{figure*}
\begin{figure*}
\psfrag{xh1}[c][c][1][0]{{$\xb$}}
\psfrag{p2/p0}[c][c][1][0]{{$P^s_2/P^s_0$}}
\psfrag{k=0.23}[c][r][1][0]{{\footnotesize $k=0.23\,{\rm Mpc}^{-1}$}}
\psfrag{k=0.55}[c][r][1][0]{{\footnotesize $k=0.55\,{\rm Mpc}^{-1}$}}
\psfrag{k=3.00}[c][r][1][0]{{\footnotesize $k=3.00\,{\rm Mpc}^{-1}$}}
\psfrag{IR}[t][b][1][0]{{\footnotesize IR}}
\psfrag{HR}[t][b][1][0]{{\footnotesize HR}}
\centering
\includegraphics[width=0.65\textwidth, angle=-90]{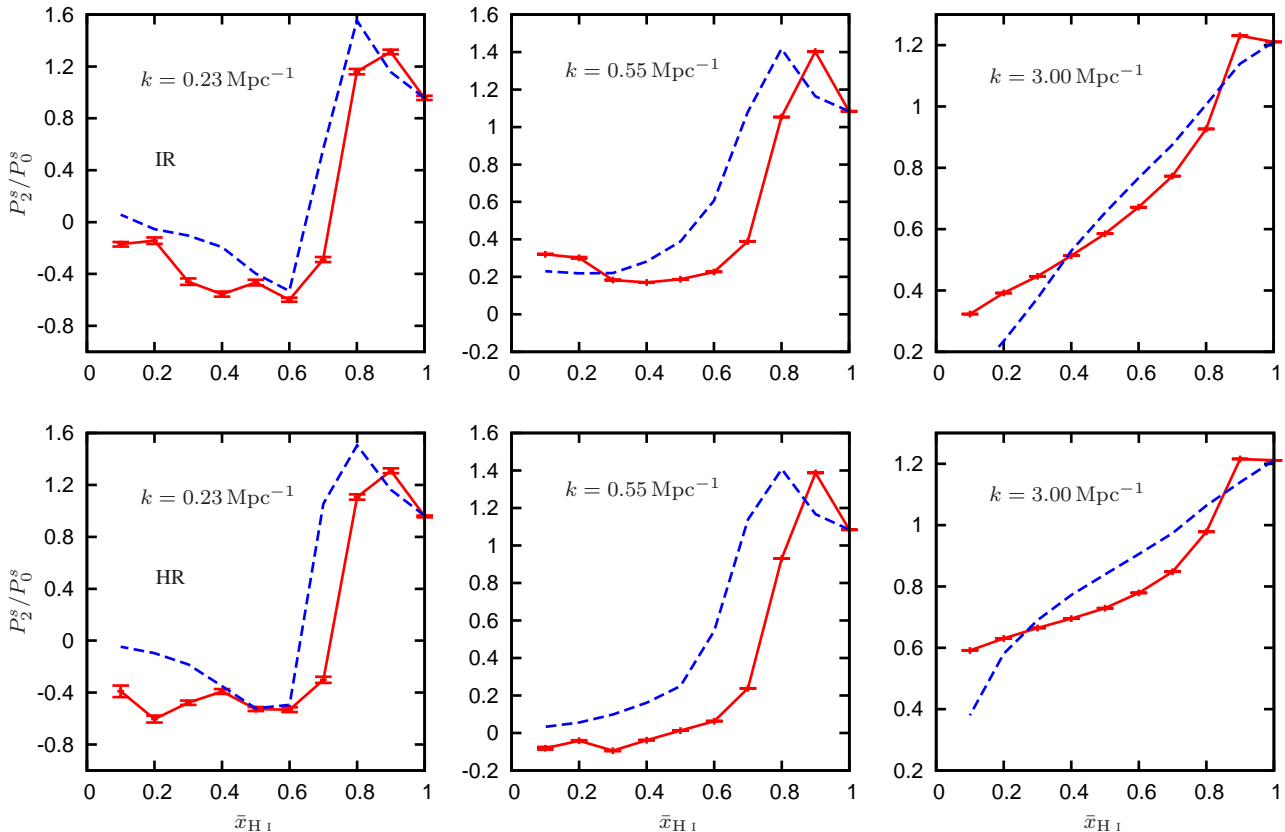}
\caption{The ratio $P^s_2/P^s_0$ as a function of $\xb$ for the three
  mentioned values of the wave number $k$. The simulation (thick solid
  line) are compared with the model prediction (dashed line). The
  error bars are estimated following the analysis discussed in
  Appendix \ref{ap:error}, excluding the effects from any specific
  telescope.}
\label{fig:p2_p0_xh1}
\end{figure*}

We have simulated the \HI distribution, in both real and redshift
space, for $\xb$ values starting from $1.0$ to $0.1$ with an interval
of $\Delta \xb = 0.1$. For each value of $\xb$ we have used 12
independent {realizations \footnote{All the results shown here (Figure
    2 to 6) are mean estimated from these 12 independent
    realizations.}}  of the simulation to estimate the mean and
$1\sigma$ error of the \HI power spectrum. The simulations are all
carried out at a fixed redshift $z = 8$. Ideally, one should adopt a
model for the evolution of $\xb$ with $z$ and simulate each neutral
fraction at the appropriate redshift. This, however, makes the results
dependent on the model for the redshift evolution of $\xb$ and also
makes the computations rather cumbersome. We have sidestepped this
issue by keeping $z$ fixed at a value $z = 8$ which is in the redshift
range that is considered favourable for observing the EoR 21-cm
signal. However due to this approach we are unable account for the
evolution for the dark matter density field as well the evolution of
the sources of ionizing photons in our simulation. We interpret the
simulations for different values of $\xb$ as representing different
stages of reionization.

Figure \ref{fig:h1_map} shows the \HI distributions at different
stages of reionization in both real and redshift space for a single
realization of the simulation. We see that the \HI distributions in
the HR and the IR models are nearly indistinguishable in the early
stages of reionization ($\xb \geq 0.7$). In the early stages
reionization proceeds inside-out in both of these models with the high
density regions ionizing first and the low density regions ionizing
later. The \HI maps of the two models become quite distinct at the
later stages of reionization ($\xb \leq 0.5$). In the IR model we find
many small isolated neutral \HI clumps distributed within the regions
which are completely ionized in the HR model. These small \HI clumps
correspond to the high density regions which become self-shielded due
to the enhancement in the recombination rate in the IR
model. Comparing the real and redshift space \HI maps we see that in
the initial stages the ionized and neutral regions respectively appear
slightly contracted and elongated along the LoS. For the HR model the
\HI distribution exhibits the same behaviour even in the late
stages. The small \HI clumps which appear in the late stages of the IR
model however, show the opposite behaviour. These neutral regions
appear contracted along the LoS. All the features mentioned above can
be understood based on the fact that overdense and underdense regions
respectively appear contracted and elongated along the LoS.

\begin{figure*}
\psfrag{xh1=0.9}[c][c][1][0]{{\scriptsize $\xb=0.9$}}
\psfrag{=0.8}[c][c][1][0]{ {\scriptsize $\,0.8$}}
\psfrag{=0.7}[c][c][1][0]{ {\scriptsize $\,0.7$}}
\psfrag{=0.6}[c][c][1][0]{ {\scriptsize $\,0.6$}}
\psfrag{=0.5}[c][c][1][0]{ {\scriptsize $\,0.5$}}
\psfrag{=0.3}[c][c][1][0]{ {\scriptsize $\,0.3$}}
\psfrag{=0.1}[c][c][1][0]{ {\scriptsize $\,0.1$}}
\psfrag{k}[c][c][1][0]{{$k\, ({\rm Mpc}^{-1})$}}
\psfrag{a}[l][l][1][0]{{$\a(k)$}}
\psfrag{r}[l][l][1][0]{{$\r(k)$}}
\psfrag{IR}[c][c][1][0]{{IR}}
\psfrag{HR}[c][c][1][0]{{HR}}
\centering
\includegraphics[width=.3\textwidth, angle=-90]{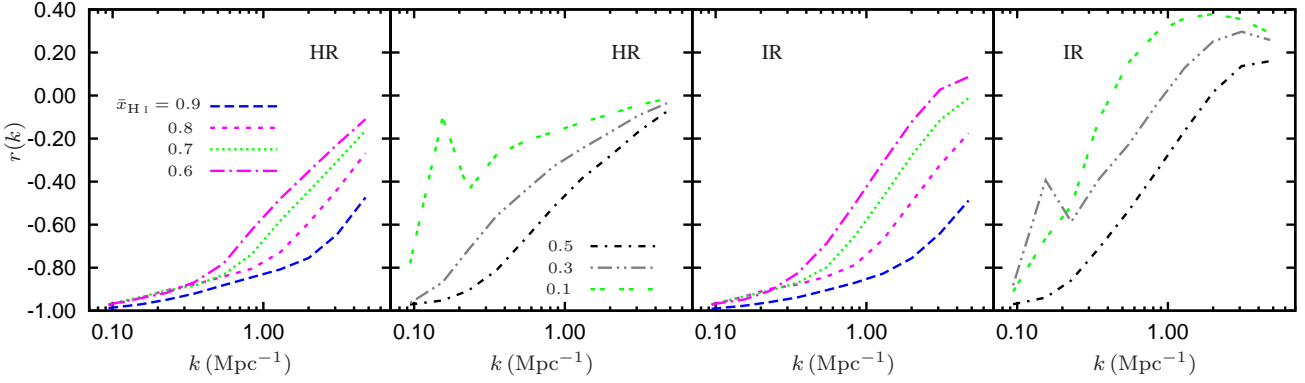}
\caption{The dimensionless cross-correlation $\r(k)$ for different
  neutral fractions and for both the models of reionization.}
\label{fig:pk_rk}
\end{figure*}

We show the ratio between the monopole of the redshift space power
spectrum with its real space counterpart in Figure
\ref{fig:pk_ratio}. This ratio ($P^s_0/P^r$) has been extensively used
in the literature ({\it e.g.}  \citealt{lidz07,mesinger11,mao12}) to
quantify the effect of redshift space distortion. The earlier works
have reported that at large length scales this ratio rises to a value
greater than $1.87$ in the early stages of reionization, and then
falls to a value slightly less than $1$ at later stages. Our results
show a similar behaviour in both the HR and IR models. We also find
that at large length scales ($k \sim 0.2\, {\rm Mpc}^{-1}$)
$P^s_0/P^r$ rises to a value $\approx 2$ at $\xb = 0.8-0.9$ and then
abruptly falls to a value slightly less than $1$ at $\xb = 0.7$. The
ratio subsequently rises gradually and approaches $1$ as $\xb$
decreases. The behaviour at intermediate length scales ($k \sim 0.5\,
{\rm Mpc}^{-1}$) is somewhat similar, except that we have a sharp peak
at $\xb = 0.9$. Also, the ratio again exceeds $1$ at $\xb \leq 0.2$
for the IR model. Several earlier works have highlighted the initial
rise in $P^s_0/P^r$ at $\xb \sim 0.8$ as a very prominent feature of
the effect of redshift space distortion on the EoR 21-cm signal. We
note, that the expected signal itself drops considerably at $\xb =
0.8$ and this is possibly not very significant from the observational
point of view. Further, the sudden rise in $P^s_0/P^r$ is possibly
because of the rapid decline in the signal itself (due to the possible
negative contribution from $P_{\Delta x}$ in eq. [\ref{eq:modelr}] and
competition between $P_{\Delta \Delta}$, $P_{\Delta x}$ and $P_{xx}$,
this is further verified later in this paper in Figure \ref{fig:pk_rk}
and \ref{fig:pk_ak}) at $\xb = 0.8$. Both $P^s_0$ and $P^r$ reduces at
this stage, however $P^r$, which appears in the denominator of this
ratio, declines much faster than $P^s_0$ (for further details see the
relevant discussion in \citealt{mao12}).

The quantity that in principle can be directly estimated from the
observational data and will quantify the strength as well as the
nature of the redshift space anisotropy is the ratio between the
quadrupole and the monopole of the redshift space power spectrum. Our
estimations for this ratio ($P^s_2/P^s_0$) from simulations has been
shown in the Figure \ref{fig:p2p0_ratio}. We find that for $\xb = 1.0$
the results from our simulations are consistent with the value $1.02$
predicted by linear theory at large length scales ($k \leq 0.5\,{\rm
  Mpc}^{-1}$). At smaller scales the ratio is larger than $1.02$
possibly because of the non-linear redshift space distortion. This
ratio increases at all length scales for $\xb = 0.9$. The ratio
increases further for $\xb = 0.8$ at large scales ($k < 0.3\,{\rm
  Mpc}^{-1}$) whereas it decreases at smaller length scales. The
behaviour of this ratio changes significantly at $\xb \leq 0.7$ where
we find that $P^s_2/P^s_0$ is negative at large scales. The value
increases steadily as we move to smaller scales, crosses zero at $k
\sim 1.0\,{\rm Mpc}^{-1}$ and is positive at smaller length
scales. The results for the HR and IR models are very similar except
that there is a bump at around $k \sim 0.5\,{\rm Mpc}^{-1}$ for the IR
model. To understand the evolution of $P^s_2/P^s_0$ with $\xb$ in more
detail we study the behaviour of this ratio as a function of $\xb$
(Figure \ref{fig:p2_p0_xh1}) at three representative $k$ modes $k =
0.23, \, 0.55\, {\rm and}\, 3.00 \,{\rm Mpc}^{-1}$ hitherto referred
to as large, intermediate and small scales. We could have, in
principle, chosen a smaller $k$ mode to illustrate the behaviour at
large scales. The errors however are rather large for $k < 0.2 \,{\rm
  Mpc}^{-1}$ and these scales do not provide a very reliable estimate
of the behaviour, and we need larger simulations to study the
behaviour at length scales as large as these. Considering the large
scale first (Figure \ref{fig:p2_p0_xh1}), the behaviour of the HR and
IR models are both quite similar. The ratio rises from $\sim 1$ at
$\xb = 1.0$ to $1.4$ at $\xb = 0.8-0.9$ and then abruptly falls to
$\approx -0.3$ at $\xb = 0.7$. The ratio remains negative with values
in the range $(-0.4) - (-0.6)$ for smaller values of $\xb$ with the
exception that it rises to $\approx -0.2$ for $\xb \leq 0.2$ in the IR
model. The behaviour at intermediate scales is very similar as that at
the large scales except that the ratio falls to a value in the range
$0.2 - 0.4$ at $\xb = 0.7$ instead of becoming negative. The ratio
declines further and is negative ($\approx -0.1$) for $\xb < 0.4$ in
the HR model. In the IR model the ratio is negative nowhere and is in
the range $0.2 - 0.4$ for $\xb < 0.7$. At small scales the ratio is
constant at $\sim 1.2$ for $\xb \geq 0.9$ where after it falls rapidly
to $0.8$ at $\xb = 0.7$ and subsequently declines gradually to $\sim
0.6$ and $\sim 0.4$ in the HR and IR models respectively.

The hexadecapole $P^s_4(k)$ measured from our simulations has very
large error bars, and consequently we have not shown these
here. Larger simulations are required for reliable estimates of
$P^s_4(k)$. However it can be noted that the linear model predicts the
ratio $P^s_4(k)/P^s_0(k)$ to be very small ($P^s_4/P^s_0 \simeq 0.12$)
even for a completely neutral IGM, when compared with the ratio
$P^s_2/P^s_0 \simeq 1.02$. Thus a successful estimation of $P^s_4(k)$
from the observational data is intrinsically a difficult task. A
recent work by \citet{shapiro13} has discussed the validity of using
the $4^{{\rm th}}$ moment of the redshift space power spectrum (in
$\mu$ decomposition technique), which is some what similar to the
hexadecapole $P^s_4(k)$, for extracting the cosmological information
from the redshifted 21 cm observations.

\section{Discussion and Conclusions}
We first compare our results with the predictions of the model
discussed in Section \ref{sec:model} for which $P^s_0$ and $P^s_2$ can
be calculated using eq.s (\ref{eq:model0}) and (\ref{eq:model2})
respectively. The model requires $P_{\Delta \Delta}(k)$ , $P_{xx}(k)$
and $P_{\Delta x}(k)$ which we have directly determined from our
simulations at different stages of reionization. Several earlier
authors \citep{lidz07,mesinger11,mao12} have noted that this model
does not correctly reproduce the real and redshift space power
spectrum $P^r$ and $P^s_0$ determined from simulations. Our results
for the $\xb$ dependence of $P^s_2/P^s_0$ shown in Figure
\ref{fig:p2_p0_xh1} confirm that the model fails to quantitatively
reproduce the results of the simulations. We however note that the
predictions of this model are qualitatively very similar to the
results of the simulations, and this provides a very useful framework
for interpreting our results. In the subsequent discussion we have
used the framework of this model to discuss and qualitatively
interpret the $\xb$ dependence of the observable quantity
$P^s_2/P^s_0$, which we have studied earlier.

\begin{figure}
\psfrag{k}[c][c][1][0]{{$k\, ({\rm Mpc}^{-1})$}}
\psfrag{a}[l][l][1][0]{{$\a(k)$}}
\psfrag{IR}[c][c][1][0]{{IR}}
\psfrag{HR}[c][c][1][0]{{HR}}
\centering
\includegraphics[width=.27\textwidth, angle=-90]{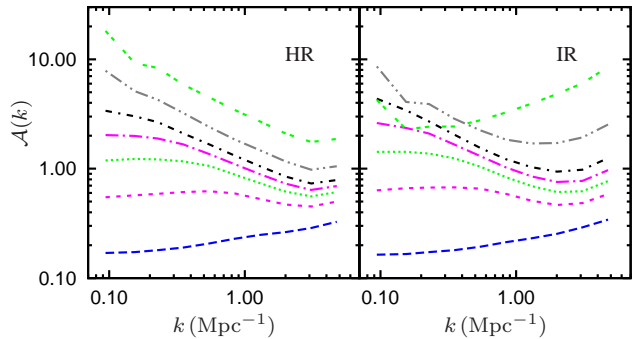}
\caption{This shows $\a(k) = \sqrt{P_{xx}(k)/P_{\Delta \Delta}(k)}$
  for different neutral fractions and for both the models of
  reionization. The line styles are the same as Figure
  \ref{fig:pk_rk}. In the left panel the bottom most curve corresponds
  to $\xb = 0.9$ and the value of $\a(k)$ increases monotonically with
  decreasing $\xb$.}
\label{fig:pk_ak}
\end{figure}

According to the linear model the quantities which will combinedly
determine the strength as well as the nature of the redshift space
distortions are the cross-correlation power spectrum ($P_{\Delta x}$)
between the matter density fluctuations ($\Delta$) and the
fluctuations in the neutral fraction ($\Delta_x$) and the power
spectrum of fluctuations in the neutral fraction field ($P_{xx}$). The
quantity $P_{xx}$ by definition will always be positive, whereas the
quantity $P_{\Delta x}$ can have both positive or negative values. A
negative/positive contribution from $P_{\Delta x}$ would represent an
anti-correlation/correlation between the $\Delta$ and $\Delta_x$
fields. To understand their relative contribution on the simulated
redshift space power spectrum we represent them in terms of two
dimensionless quantities 
\begin{equation}
  \a(k) = \sqrt{P_{xx}(k)/P_{\Delta\Delta}(k)} \,,
\label{eq:ak}
\end{equation}
 and 
\begin{equation}
\r(k) = P_{\Delta x}(k)/\sqrt{P_{xx}(k) P_{\Delta \Delta}(k)}\,,
\label{eq:rk}
\end{equation}
from our simulations. The quantity $\a(k)$ defines the relative
amplitude of $P_{xx}$ with respect to $P_{\Delta \Delta}$, whereas
$\r(k)$ determines the strength of cross-correlation between $\Delta$
and $\Delta_x$. Figure \ref{fig:pk_rk} and \ref{fig:pk_ak} show
$\r(k)$ and $\a(k)$ respectively at different stages of
reionization. Considering the large scales first we find that $\r
\simeq -1$ at all stages of reionization in both the HR and IR
models. This indicate that at large scales the distribution of neutral
fraction is anti-correlated with the matter distribution throughout
reionization, this being a consequence of the fact that the high
density regions are the locations where we expect to find the sources
that drives reionization thus they are expected to get ionized first
and only the low density regions remain neutral. Note that the small
neutral clumps produced at the later stages in the IR model do not
affect this behaviour seen at large scales. At all scales we find
$\a(k)$ to monotonically increase with decreasing $\xb$. The observed
sharp peak of the ratio $P^s_2/P^s_0$ (left panels of Figure
\ref{fig:p2_p0_xh1}) at the early stages of reionization ($\xb \simeq
0.9$) at large and intermediate scales thus represent the fact that at
this stage the contribution from $P_{xx}$ becomes comparable to
$P_{\Delta \Delta}$ and due to a complete anti-correlation between
$\Delta$ and $\Delta_x$ ({\it i.e.}  $\r \simeq -1$) the actual signal
($P^s_0$) becomes very low, which appears at the denominator of this
ratio. This peak thus should not be misinterpreted as a signature of
redshift space distortion. At the later stages of reionization ($0.1 <
\xb \leq 0.8$ for the HR and $0.3 < \xb \leq 0.8$ for the IR model)
the cross correlation still remains $\r \simeq -1$ whereas
$P_{xx}>P_{\Delta \Delta}$ which leads to a negative value of
$P^s_2/P^s_0$ in both models of reionization. These results points
towards an inside-out reionization at large scales. We observe that
the presence of small neutral clumps, produced at the later stages in
the IR model, does not contribute significantly at the large scale
power spectrum. Thus a negative value of $P^s_2/P^s_0$ of the 21-cm
power spectrum itself will be a direct evidence that the reionization
has happened and it is inside-out at large scale.

The fluctuations in the neutral fraction $\Delta_x(k)$ and the matter
$\Delta(k)$ are not perfectly anti-correlated ($\r > -1$) at
intermediate and small scales. The value of $\r$ increases as we go to
smaller scales, and the behaviour is similar in the HR and IR models
at the early stages of reionization ($\xb \geq 0.8$). We see
differences between the HR and IR model at intermediate and small
scales at the later stages of reionization. The values of $\r$ are
larger in the IR model in comparison to the HR model, and $\r$ becomes
positive at small scales in the later stages in the IR model whereas
it is nowhere positive in the HR model. This difference is the outcome
of the small \HI clumps seen in Figure \ref{fig:h1_map} at the later
stages in the IR model. The signature of this difference is also
visible in the observed power spectrum ($P^s_2/P^s_0$) at the
intermediate scales (middle panels of Figure \ref{fig:p2_p0_xh1}). At
intermediate scales the ratio $P^s_2/P^s_0$ becomes negative during
the later stages of reionization ($\xb \leq 0.6$) in the HR model,
however in IR model it never becomes negative at these scales and more
or less maintains a constant positive value of $\approx 0.2$ at the
later stages of reionization. 

Thus the two major signatures from the redshift space 21-cm signal
that can be used as the evidence of reionization as well as a
characterization for the redshift space anisotropy are the following
---
\begin{itemize}
\item A negative value of the ratio $P^s_2/P^s_0$ at large scales during the
  intermediate and late stages of reionization.
\item The ratio $P^s_2/P^s_0$ stays negative even at intermediate
  scales for a completely inside-out reionization whereas it becomes
  positive for a partially outside-in reionization.
\end{itemize} 
The main point of concern here is how unambiguously it will be
possible to detect these signatures, or in other words what level of
sensitivity our measurements will require to detect these signatures.
To get an idea of the accuracy level of the measurements required and
to find out the possible effects of uncertainty, we have done a rough
error analysis for estimations of the ratios of various angular
multipoles. This analysis has been discussed in detail in Appendix
\ref{ap:error}. Following this analysis (eq. [\ref{eq:delR}]) we have
estimated the possible errors in our estimation of the ratio $P_2/P_0$
and shown them as error bars in Figure \ref{fig:p2_p0_xh1}. However
while estimating these errors we have not considered the possible
uncertainties arising from the system noise of a specific observation
with a telescope and also replaced the cosmic variance in
eq. (\ref{eq:delR}) with the sample variance of the zeroth moment of
the power spectrum. We plan to include the effect of system noise and
various other observation specific effects in our future work. From
our analysis in Appendix \ref{ap:error} and results in Figure
\ref{fig:p2_p0_xh1} it is evident that it will be possible to
unambiguously detect both these signatures of reionization at large
and intermediate scales, to which the present ({\it e.g.} LOFAR, see
\citet{jensen13} for more details) and upcoming telescopes will be
sensitive, if the system noise can be suppressed to a sufficient
level.

Another alternative approach to quantify the strength and the nature
of redshift space distortion from the \HI power spectrum is by
decomposing the redshift space power spectrum in the various
coefficients of the powers of $\mu$. These estimated coefficients can
be then interpreted following a linear \citep{barkana05} or a
quasi-linear \citep{mao12} model. However in this method the estimated
coefficients of the powers of $\mu$ will not be completely independent
of each other and the correlation between them (or the leakage of
power from one component to another) may give rise to wrong
interpretations as observed in a recent work by \citet{jensen13}. In
comparison to this method the decomposition of the anisotropy due to
peculiar velocities using Legendre polynomials, is a representation in
orthonormal basis therefore the various angular multipoles estimated
through this method will be independent of each other (as discussed in
Appendix \ref{ap:error}). Thus one does not have to be too concerned
about the leakage of power between different multipoles in this
method.

It is anticipated that the initial observational attempts ({\it e.g.}
GMRT, LOFAR and MWA) will probe the EoR 21-cm signal only at large
scales. The multipole moments of the power spectrum of the measured
signal hold the key to quantify and interpret it.  We have considered
the ratio between the quadrupole and monopole moments which is capable
of quantifying the strength of the signal and the anisotropy which
arises due to the peculiar velocities. Our simulations indicate that
the prospects of detecting the EoR 21-cm signal are most favourable
when the mean neutral fraction is in the range $\xb=0.4-0.5$. In this
range the signal, we see, is characterized by two main features for
both the HR and IR models of reionization. First, the monopole moment
of the redshift space \HI power spectrum $P^s_0(k)$ is nearly equal to
the real space \HI power spectrum $P^r(k)$, and both of these are
comparable to the real space matter power spectrum.  Second, the
quadrupole moment of the redshift space \HI power spectrum $P^s_2(k)$
is negative with a value which is $-0.5$ times $P^s_0(k)$.  This is in
contrast to the value $1.02$ predicted by linear theory for the ratio
$P^s_2/P^s_0$ of the matter power spectrum. We also observe that it
would be possible to distinguish between the inside-out and the
outside-in reionization scenarios from the nature of the ratio between
the quadrupole and monopole moments at the intermediate length scales
($k \simeq 0.5 \,{\rm Mpc}^{-1}$). This particular signature may help
in ruling out the extremely outside-in reionization models using
future observations.
 
\section{acknowledgments}
Suman Majumdar would like to thank Prof. Sugata Pratik Khastgir for
useful discussions and pointing out some errors in the initial version
of the manuscript. Suman Majumdar would like to acknowledge Council of
Scientific and Industrial Research (CSIR), India for providing
financial assistance through a senior research fellowship (File
No. 9/81 (1099)/10-EMR-I).

\appendix

\section{Quantifying the possible uncertainties in the estimation of
  angular moments of the 21-cm power spectrum}
\label{ap:error}
In this section we provide a rough analysis for quantifying the
possible uncertenties in the estimation of angular multipoles of power
spectrum and their ratios. We consider redshifted 21-cm observations
that cover a 3-D volume $V$. In the continuum limit, the number of
independent Fourier modes $d N$ corresponding to $d^3 k$ is
\begin{equation}
dN= V \frac{d^3 k}{(2 \pi)^3}
\end{equation}

We consider $\P(\k)$ which is an unbiased estimator of the power
spectrum at the Fourier mode $\k$. This has the properties that
\begin{equation}
\langle \P(\k) \rangle = P(\k)
\label{eq:est1}
\end{equation}
and 
\begin{equation}
\langle [\Delta \P(\k)] [\Delta \P(\k^{'})] \rangle = \frac{(2
  \pi)^3}{V} \delta(\k-\k^{'}) \sigma^2(\k) 
\label{eq:est2}
\end{equation}
where the latter equation essentially tells us that the estimates at
the different Fourier modes are independent. Also the variance of
power spectrum estimator has two independent contributions 
\begin{equation}
\sigma^2 (\k)=[P(\k)+\sigma_N^2]^2
\label{eq:est3}
\end{equation}
which arise from the cosmic variance $[P(\k)]$  and the system noise 
$[\sigma_N^2]$ respectively. The system noise is inherent to the
observations and it depends on the observing time, the telescope,
etc. 

We use this to define estimators for the multipole moments of the
power spectrum 
\begin{equation}
\P_{n}(k_i) = \frac{ \int d^3 k \, \Pl_n(\mu) \P(\k)}{
  \int d^3 k \, \Pl_n^2(\mu) }
\label{eq:mom1}
\end{equation}
where $\Pl_n(\mu)$ is the Legendre polynomial of order $n$, $\mu=\k
\cdot {\bf n}/k$ is the cosine of the angle between $\k$ and the line
of sight ${\bf n}$, and the $d^3 \, k$ integral is over a spherical
shell of radius $k_i$ and width $\Delta k_i$. 

We have 
\begin{equation}
\langle  \P_n(k_i)  \rangle = P_n(k_i) \,.
\end{equation}
We assume that the bins have no overlap, whereby it is obvious that
the estimators $P_n(k_i)$ and $P_m(k_j)$ in two different bins $(k_i
\neq k_j)$ are uncorrelated.  We now calculate the covariance between
the multipoles in the same bin
\begin{align}
  \langle [\Delta \P_n(k_i)] [\Delta \P_m(k_i)] \rangle =& \left
    \langle \frac{ \int d^3 k_1 \, \Pl_n(\mu_1) [\Delta \P(\k_1)]}{
      \int d^3 k_1 \, \Pl_n^2(\mu_1) } \right.\nonumber\\
  &\left.\frac{ \int d^3 k_2 \, \Pl_m(\mu_2) [\Delta \P(\k_2)]}{ \int
      d^3 k_2 \, \Pl_m^2(\mu_2) } \right \rangle
\end{align}
which using eq. (\ref{eq:est2}) gives
\begin{align}
  \langle [\Delta \P_n(k_i)]& [\Delta \P_m(k_i)] \rangle =\nonumber\\
&\frac{ (2 \pi)^3 V^{-1} \int d^3 k_1 \, \Pl_n(\mu_1) \Pl_m(\mu_1)
    \sigma^2(\k)} {   \int d^3 k_1 \, \Pl_n^2(\mu_1)   \int d^3 k_2 \,
    \Pl_m^2(\mu_2) } 
\end{align}
This can be further simplified using
\begin{align}
  \int d^3 k \Pl_n(\mu_1) &\Pl_m(\mu_1) \sigma^2(\k)= \nonumber\\ & 2 \pi k_i^2 \,
  \Delta k_i \, \sigma^2(k_i) 
  \int_{-1}^{1} d \mu_1 \, \Pl_n(\mu_1) \Pl_m(\mu_1)
\end{align}
and 
\begin{equation}
\int_{-1}^{1} d \mu_1 \,  \Pl_n(\mu_1) \Pl_m(\mu_1)  = \frac{2
  \delta_{n,m}}{2 n +1 }
\end{equation} 
whereby 
\begin{equation}
\langle [\Delta \P_n(k_i)] [\Delta \P_m(k_i)] \rangle
= \delta_{m,n} \frac{2 \pi^2 (2 n+1) \sigma^2(k_i)}{V k_i^2 \, \Delta
  k_i} 
\end{equation}
We see that the errors in the different multipole moments are
uncorrelated.  For the monopole 
\begin{equation}
\langle [\Delta \P_0(k_i)]^2  \rangle
=  \frac{2 \pi^2 [P(\k)+\sigma_N^2]^2}{V k_i^2 \, \Delta
  k_i} 
\end{equation}
and 
\begin{equation}
\langle [\Delta \P_n(k_i)]^2  \rangle = (2 n +1)
\langle [\Delta \P_0(k_i)]^2  \rangle  
\end{equation}
for the higher multipoles. 
 
We next consider the ratio of the multipoles  
\begin{equation}
\Rh_n(k_i)=\frac{\P_n(k_i)}{\P_0(k_i)}
\end{equation}
We calculate the variance of $\Rh_n(k_i)$ using  
\begin{equation}
\Delta \Rh_n(k_i) = \frac{[\Delta \P_n(k_i)]}{P_0(k_i)}
- \frac{P_n(k_i)}{P_0(k_i)} \frac{[\Delta \P_0(k_i)]}{P_0(k_i)}
\end{equation}
whereby 
\begin{equation}
\langle [\Delta \Rh_n(k_i)]^2 \rangle  = \frac{\langle [\Delta
    \P_n(k_i)]^2 \rangle }{P_0^2(k_i)} 
+ \frac{P_n^2(k_i)}{P_0^2(k_i)} \frac{\langle [\Delta \P_0(k_i)]^2 \rangle
}{P_0^2(k_i)} 
\end{equation}
which gives 
\begin{equation}
\langle [\Delta \Rh_n(k_i)]^2 \rangle  = \left[ (2n  + 1) + \R_n^2(k_i)
  \right]    \frac{\langle [\Delta \P_0(k_i)]^2 \rangle}{P_0^2(k_i)}  
\end{equation}
In summary we have a relation between $\delta \R_n$ which is the
error  in the ratio $\R_n=P_n/P_0$  and the fractional error $\delta
P_o/P_o$ of  the   monopole. 
\begin{equation}
\delta \R_n = \sqrt{2 n + 1 + \R^2_n} \ \left(\frac{\delta P_0}{P_0}
\right) 
\label{eq:delR}
\end{equation}
This provides a rough estimation of possible errors that will be
present in the estimations of the ratio of angular moments of 21-cm
power spectrum. We plan to take up a more detailed uncertainty
analysis in our future work.


\begin{thebibliography}{99}
\bibitem[\protect\citeauthoryear{Alvarez et al.}{2006}]{alvarez06}
  Alvarez, M.~A., Shapiro, P.~R., Ahn, K., \& Iliev, I.~T.\ 2006,
  \apjl, 644, L101


\bibitem[\protect\citeauthoryear{Alvarez et al.}{2009}]{alvarez09}
  Alvarez, M.~A., Busha, M., Abel, T., \& Wechsler, R.~H.\ 2009,
  \apjl, 703, L167

\bibitem[\protect\citeauthoryear{Baek et al.}{2009}]{baek09} Baek, S.,
  Di Matteo, P., Semelin, B., Combes, F., \& Revaz, Y.\ 2009, \aap,
  495, 389

\bibitem[\protect\citeauthoryear{Barkana \& Loeb}{2005}]{barkana05}
  Barkana, R., \& Loeb, A.\ 2005, \apjl, 624, L65

\bibitem[\protect\citeauthoryear{Becker et al.}{2001}]{becker01}
  Becker, R.~H., Fan, X., White, R.~L., et al.\ 2001, \aj, 122, 2850

\bibitem[\protect\citeauthoryear{Bharadwaj et al.}{2001}]{bharadwaj01}
  Bharadwaj, S., Nath, B.~B., \& Sethi, S.~K.\ 2001, Journal of
  Astrophysics and Astronomy, 22, 21

\bibitem[\protect\citeauthoryear{Bharadwaj \& Ali}{2004}]{bharadwaj04}
  Bharadwaj, S., \& Ali, S.~S.\ 2004, \mnras, 352, 142

\bibitem[\protect\citeauthoryear{Bharadwaj \& Ali}{2005}]{bharadwaj05}
  Bharadwaj, S., \& Ali, S.~S.\ 2005, \mnras, 356, 1519

\bibitem[\protect\citeauthoryear{Bouwens et al.}{2005}]{bouwens05}
  Bouwens, R.~J., Illingworth, G.~D., Thompson, R.~I., \& Franx,
  M.\ 2005, \apjl, 624, L5

\bibitem[\protect\citeauthoryear{Cole et al.}{1995}]{cole95} Cole, S.,
  Fisher, K.~B., \& Weinberg, D.~H.\ 1995, \mnras, 275, 515

\bibitem[\protect\citeauthoryear{{Choudhury} \&
    {Ferrara}}{2006}]{choudhury06} {Choudhury}, T.~R., {Ferrara}, A.,
  2006, Cosmic Polarization, Editor - R. Fabbri(Research Signpost),
  p. 205, arXiv:astro-ph/0603149

\bibitem[\protect\citeauthoryear{{Choudhury} \&
    {Ferrara}}{2006}]{choudhury06a} Choudhury, T.~R., \& Ferrara,
  A.\ 2006, Albert Einstein Century International Conference, 861, 835

\bibitem[\protect\citeauthoryear{Choudhury}{2009}]{choudhury09a}
Choudhury, T.~R.,  2009, Current Science, 97, 6, 841

\bibitem[\protect\citeauthoryear{Choudhury et al.}{2009}]{choudhury09} Choudhury, T.~R., Haehnelt, M.~G., \&
  Regan, J.\ 2009, \mnras, 394, 960

\bibitem[\protect\citeauthoryear{Ciardi et al.}{2001}]{ciardi01}
  Ciardi, B., Ferrara, A., Marri, S., \& Raimondo, G.\ 2001, \mnras,
  324, 381

\bibitem[\protect\citeauthoryear{Davis et al.}{1985}]{davis} 
Davis,  M., Efstathiou, G., Frenk, C.~S., \& White, S.~D.~M.
\ 1985, \apj, 292, 371 

\bibitem[\protect\citeauthoryear{Fan et al.}{2003}]{fan03} Fan, X.,
  Strauss, M.~A., Schneider, D.~P., et al.\ 2003, \aj, 125, 1649

\bibitem[\protect\citeauthoryear{Fan et al.}{2006}]{fan06} Fan, X.,
  Carilli, C.~L., \& Keating, B.\ 2006, \araa, 44, 415

\bibitem[\protect\citeauthoryear{Furlanetto et
    al.}{2004}]{furlanetto04} {Furlanetto}, S.~R., {Zaldarriaga},
  M. \& {Hernquist}, L. 2004, \apj, 613, 1

\bibitem[\protect\citeauthoryear{Geil \& Wyithe}{2008}]{geil08a} Geil,
  P.~M., \& Wyithe, J.~S.~B.\ 2008, \mnras, 386, 1683

\bibitem[\protect\citeauthoryear{Gnedin}{2000}]{gnedin00} Gnedin,
  N.~Y.\ 2000, \apj, 535, 530

\bibitem[\protect\citeauthoryear{Goto et al.}{2011}]{goto11} Goto, T.,
  Utsumi, Y., Hattori, T., Miyazaki, S., \& Yamauchi, C.\ 2011,
  \mnras, 415, L1

\bibitem[\protect\citeauthoryear{Hamilton}{1992}]{hamilton92}
  Hamilton, A.~J.~S.\ 1992, \apjl, 385, L5

\bibitem[\protect\citeauthoryear{Hamilton}{1998}]{hamilton98}
  Hamilton, A.~J.~S.\ 1998, The Evolving Universe, 231, 185

\bibitem[\protect\citeauthoryear{Iliev et al.}{2006}]{iliev06} Iliev,
  I.~T., Pen, U.-L., Richard Bond, J., Mellema, G., \& Shapiro,
  P.~R.\ 2006, New Astronomy Reviews, 50, 909

\bibitem[\protect\citeauthoryear{Iliev et al.}{2008}]{iliev08} Iliev,
  I.~T., Mellema, G., Pen, U.-L., Bond, J.~R., \& Shapiro,
  P.~R.\ 2008, \mnras, 384, 863

\bibitem[\protect\citeauthoryear{Jarosik et al.}{2011}]{jarosik11}
  Jarosik, N., Bennett, C.~L., Dunkley, J., et al.\ 2011, \apjs, 192,
  14

\bibitem[\protect\citeauthoryear{Jensen et al.}{2013}]{jensen13}
  Jensen, H., Datta, K.~K., Mellema, G., et al.\ 2013, arXiv:1303.5627

\bibitem[\protect\citeauthoryear{Kaiser}{1987}]{kaiser87} Kaiser,
  N.\ 1987, \mnras, 227, 1

\bibitem[\protect\citeauthoryear{Komatsu et al.}{2011}]{komatsu11}
  Komatsu, E., Smith, K.~M., Dunkley, J., et al.\ 2011, \apjs, 192, 18

\bibitem[\protect\citeauthoryear{Larson et al.}{2011}]{larson11}
  Larson, D., Dunkley, J., Hinshaw, G., et al.\ 2011, \apjs, 192, 16

\bibitem[\protect\citeauthoryear{Lidz et al.}{2007}]{lidz07} Lidz, A.,
  Zahn, O., McQuinn, M., et al.\ 2007, \apj, 659, 865

\bibitem[\protect\citeauthoryear{Lidz et al.}{2009}]{lidz09} Lidz, A.,
  Zahn, O., Furlanetto, S.~R., et al.\ 2009, \apj, 690, 252

\bibitem[\protect\citeauthoryear{Mao et al.}{2012}]{mao12} Mao, Y.,
  Shapiro, P.~R., Mellema, G., et al.\ 2012, \mnras, 422, 926

\bibitem[\protect\citeauthoryear{Maselli et al.}{2003}]{maselli03}
  Maselli, A., Ferrara, A., \& Ciardi, B.\ 2003, \mnras, 345, 379

\bibitem[\protect\citeauthoryear{McQuinn et al.}{2007}]{mcquinn07}
  McQuinn, M., Lidz, A., Zahn, O., et al.\ 2007, \mnras, 377, 1043

\bibitem[\protect\citeauthoryear{Mellema et al.}{2006}]{mellema06}
  Mellema, G., Iliev, I.~T., Pen, U.-L., \& Shapiro, P.~R.\ 2006,
  \mnras, 372, 679

\bibitem[\protect\citeauthoryear{Mesinger \&
    Furlanetto}{2007}]{mesinger07} Mesinger, A., \& Furlanetto,
  S.\ 2007, \apj, 669, 663

\bibitem[\protect\citeauthoryear{Mesinger et al.}{2011}]{mesinger11}
  Mesinger, A., Furlanetto, S., \& Cen, R.\ 2011, \mnras, 411, 955

\bibitem[\protect\citeauthoryear{Mitra et al.}{2011}]{mitra11} Mitra,
  S., Choudhury, T.~R., \& Ferrara, A.\ 2011, \mnras, 413, 1569

\bibitem[\protect\citeauthoryear{Page et al.}{2007}]{page07} Page, L.,
  Hinshaw, G., Komatsu, E., et al.\ 2007, \apjs, 170, 335

\bibitem[\protect\citeauthoryear{Razoumov et al.}{2002}]{razoumov02}
  Razoumov, A.~O., Norman, M.~L., Abel, T., \& Scott, D.\ 2002, \apj,
  572, 695

\bibitem[\protect\citeauthoryear{Ricotti et al.}{2002}]{ricotti02}
  Ricotti, M., Gnedin, N.~Y., \& Shull, J.~M.\ 2002, \apj, 575, 33

\bibitem[\protect\citeauthoryear{Santos et al.}{2010}]{santos10}
  Santos, M.~G., Ferramacho, L., Silva, M.~B., Amblard, A., \& Cooray,
  A.\ 2010, \mnras, 406, 2421

\bibitem[\protect\citeauthoryear{Semelin et al.}{2007}]{semelin07}
  Semelin, B., Combes, F., \& Baek, S.\ 2007, \aap, 474, 365

\bibitem[\protect\citeauthoryear{Shapiro et al.}{2008}]{shapiro08}
  Shapiro, P.~R., Iliev, I.~T., Mellema, G., Pen, U.-L., \& Merz,
  H.\ 2008, The Evolution of Galaxies Through the Neutral Hydrogen
  Window, 1035, 68

\bibitem[\protect\citeauthoryear{Shapiro et al.}{2013}]{shapiro13}
  Shapiro, P.~R., Mao, Y., Iliev, I.~T., et al.\ 2013, Physical Review
  Letters, 110, 151301

\bibitem[\protect\citeauthoryear{Shin et al.}{2008}]{shin08} Shin,
  M.-S., Trac, H., \& Cen, R.\ 2008, \apj, 681, 756


\bibitem[\protect\citeauthoryear{Sokasian et al.}{2003}]{sokasian03}
  Sokasian, A., Abel, T., Hernquist, L., \& Springel, V.\ 2003,
  \mnras, 344, 607

\bibitem[\protect\citeauthoryear{Spergel et al.}{2003}]{spergel03}
  Spergel, D.~N., Verde, L., Peiris, H.~V., et al.\ 2003, \apjs, 148,
  175

\bibitem[\protect\citeauthoryear{Stiavelli et al.}{2004}]{stiavelli04}
  Stiavelli, M., Fall, S.~M., \& Panagia, N.\ 2004, \apjl, 610, L1

\bibitem [\protect\citeauthoryear{Swarup et al.}{1991}]{swarup}  
Swarup G., Ananthakrishnan S., Kapahi V.K., Rao A.P., Subramanya
C.R., Kulkarni V.K.,1991 Curr.Sci.,60,95

\bibitem[\protect\citeauthoryear{Thomas et al.}{2009}]{thomas09}
  Thomas, R.~M., Zaroubi, S., Ciardi, B., et al.\ 2009, \mnras, 393,
  32

\bibitem[\protect\citeauthoryear{Trac \& Cen}{2007}]{trac07} Trac, H.,
  \& Cen, R.\ 2007, \apj, 671, 1

\bibitem[\protect\citeauthoryear{Wang \& Hu}{2006}]{wang06} Wang, X.,
  \& Hu, W.\ 2006, \apj, 643, 585

\bibitem[\protect\citeauthoryear{White et al.}{2003}]{white03} White,
  R.~L., Becker, R.~H., Fan, X., \& Strauss, M.~A.\ 2003, \aj, 126, 1

\bibitem[\protect\citeauthoryear{Willott et al.}{2007}]{willott07}
  Willott, C.~J., Delorme, P., Omont, A., et al.\ 2007, \aj, 134, 2435

\bibitem[\protect\citeauthoryear{Yan \& Windhorst}{2004}]{yan04} Yan,
  H., \& Windhorst, R.~A.\ 2004, \apjl, 600, L1

\bibitem[\protect\citeauthoryear{Zahn et al.}{2007}]{zahn07} Zahn, O.,
  Lidz, A., McQuinn, M., et al.\ 2007, \apj, 654, 12


\bibitem[\protect\citeauthoryear{Zahn et al.}{2011}]{zahn11} Zahn, O.,
  Mesinger, A., McQuinn, M., et al.\ 2011, \mnras, 414, 727

\end{thebibliography}
\end{document}